\documentclass[%
 reprint,
superscriptaddress,
 amsmath,amssymb,
 aps,
 pra,
]{revtex4-2}

\usepackage{graphicx}
\usepackage{dcolumn}
\usepackage{bm}
\usepackage{xcolor}
\usepackage[normalem]{ulem}

\begin{document}

\preprint{APS/123-QED}

\title{Intensity correlations in transmission and four-wave mixing signals intermediated by hot rubidium atoms}

  \author{A. A. C. de Almeida}
  \email{alexandre.cavalcantialmeid@ufpe.br}
 \affiliation{Departamento de F\'{i}sica, Universidade Federal de Pernambuco, 50670-901, Recife, Pernambuco, Brazil}
 \author{A. S. Alvarez}
 \affiliation{Departamento de F\'{i}sica, Universidade Federal de Pernambuco, 50670-901, Recife, Pernambuco, Brazil}
\author{N. R. de Melo}
\affiliation{Departamento de F\'{i}sica, Universidade Federal Rural de Pernambuco, 52171-050, Recife, Pernambuco, Brazil}
\author{M. R. L. da Motta}
\affiliation{Departamento de F\'{i}sica, Universidade Federal de Pernambuco, 50670-901, Recife, Pernambuco, Brazil}
\author{M. P. M. de Souza}
\affiliation{Departamento de F\'{\i}sica, Universidade Federal de Rond\^onia, 76900-726, Ji-Paran\'a, Rond\^onia, Brazil}
\author{S. S. Vianna}%
\email{sandra.vianna@ufpe.br}
\affiliation{Departamento de F\'{i}sica, Universidade Federal de Pernambuco, 50670-901, Recife, Pernambuco, Brazil}

\date{\today}

\begin{abstract}
We investigate the influence of the distribution of atom velocities in a hot rubidium sample on the correlation between field-intensity fluctuations of two independently generated four-wave mixing signals and between the transmission signals. The nonlinear process is driven by a single cw laser in a pure two-level system due to the forward geometry with circular and parallel polarization of the input fields. The intensity cross-correlations of the four-wave mixing signals and the transmission signals present an oscillatory behavior with a clear dependence on the power of the incident fields, which indicates a connection with Rabi oscillations. A two-level theoretical model using stochastic differential equations to account for the mechanism of conversion of phase noise into amplitude noise shows good agreement with our experimental results. Moreover, we show how the response of the system is affected by the different atomic velocity groups. 
\end{abstract}

\maketitle


\section{\label{sec:level1}Introduction}

The study of nonlinear interactions of light with an atomic sample has been fundamental to understanding several problems. In particular, it is well known that some features of the incident beams are transferred to the atoms during the interaction process, which consequently modifies the output light, such as the transmissions and four-wave mixing signals. Especially in the case of phase fluctuations, the resonant interaction with the atoms shifts them into intensity correlations between the transmission beams that can be directly measured \cite{ariunbold2010}.

This process, in which stochastic phase fluctuations, an intrinsic characteristic of cw diode laser radiation, are converted to amplitude modulation during the resonant interaction with an atomic medium, had its first theoretical framework given by Walser and Zoller \cite{walser1994}. Since then, this process has been investigated and explored in different ways in recent decades. As a spectroscopic technique, it provides information about the structure of the energy levels of the resonant medium, according to the pioneering experimental work of Yabuzaki et al. \cite{yabusaki1991}. Or as a tool to investigate the dynamical response of quantum systems to randomly fluctuating fields. 

Many interesting results have been produced from the study of these fluctuations in the light-matter interaction, such as the study of correlations and anticorrelations in electromagnetically induced transparency \cite{martinelli2004, cruz2007, xiao2009, florez2013}, the control of intensity noise correlations and the squeezing of four-wave-mixing processes via polarization \cite{li2016}, and the generation of correlated and anticorrelated fields via atomic spin coherence \cite{yang2012}. Part of these works are performed in the frequency domain, where the spectral decomposition of the noisy atomic response can reveal resonant spectral features, and part in the time domain. 

In this work, we focus our analysis on the time domain and study the behavior of the second-order correlation function between the intensity fluctuations of two transmission signals and two forward four-wave mixing signals generated in hot rubidium (Rb) vapor. This study differs from a previous study conducted on cold atoms \cite{almeida2023} since we discuss here the influence of the distribution of atom velocities on the observed correlation functions. Furthermore, in our theoretical treatment, we now separate the response of the FWM signals from the transmission signals, which allows us to discuss the actual contribution of the stochastic phase fluctuation to each process. For this, we consider that the phase fluctuation in the incident fields satisfies a Wiener process. Then we solve the stochastic differential system of Bloch equations using Itô’s calculus taking into account the influence of the velocity distribution of the atoms in the heated vapor.

We observed peak correlation values over 0.9 at zero delays ($\tau=0$) in transmission and four-wave mixing signals. For nonzero delays, we can have significant anticorrelation values depending on the intensity of the incident beams, indicating an oscillatory behavior compatible with Rabi oscillations \cite{papoyan2021} in the correlation functions. Similar oscillatory behavior was also observed in a magneto-optical trap of Rb atoms \cite{almeida2023}, together with high intensity correlations when incident beams have circular parallel polarizations.  In this case, the oscillatory pattern is a signature of Rabi oscillations due to the interaction with the cold atoms, which have only zero velocity.
However, as discussed in Ref. \cite{papoyan2021}, observing Rabi oscillations in a thermal vapor with a cw excitation requires some trigger mechanism or process that can lead to abrupt excitation of the atoms. In this sense, our theoretical model takes into account the integration over the Maxwell-Boltzmann velocity distribution, allowing us to show that an effective pulsed excitation regime can be produced by the flight of unexcited atoms into the region of the laser beam.

Furthermore, in the case of four-wave mixing signals, this oscillatory behavior in the second-order correlation function appears more evident. These features point out two interesting characteristics of the nonlinear process: (i) the FWM signal also contains spectral information on the atom-field interaction, and (ii) it shows that there is a dominant contribution from a specific group of atoms. 

In the following, in Sec. \ref{sec:level2} we detail
the experiment and all the experimental results. In particular,
we show the time series of all four signals, transmission and FWM, and the corresponding second-order correlation functions. Moreover, we
demonstrate an oscillatory behavior in the correlations with a clear dependence on the power of the
incident fields, which indicates a connection with Rabi oscillations. Section \ref{sec:level3} focuses on developing a simple theoretical model that allows us to separate the response of the FWM signals from the transmission signals. The theoretical results obtained from the stochastic differential equations are presented in Sec. \ref{sec:level4}, in which we detail how the system response is influenced by the different groups of atomic velocity. We
conclude by summarizing the relevant achievements of this
work in Sec. \ref{sec:level5}.

\section{\label{sec:level2}Experimental setup and results}

In the experiment, we used a single cw laser to generate two input laser beams labeled $E_a$ and $E_b$ with wave-vectors $\vec{k}_a$ and $\vec{k}_b$, as presented in Fig. 1. These two beams with circular and parallel polarization interact with a heated sample of Rb atoms contained in a glass cell wrapped in $\mu$ metal for magnetic shielding. For the temperatures with which we typically work, $T\approx60-70\,^\circ\mathrm{C}$, we expect an atomic density of the order of $10^{11}\,\mathrm{atoms}/\mathrm{cm}^3$. 

\begin{figure}
\centering
\includegraphics[width=1\linewidth]{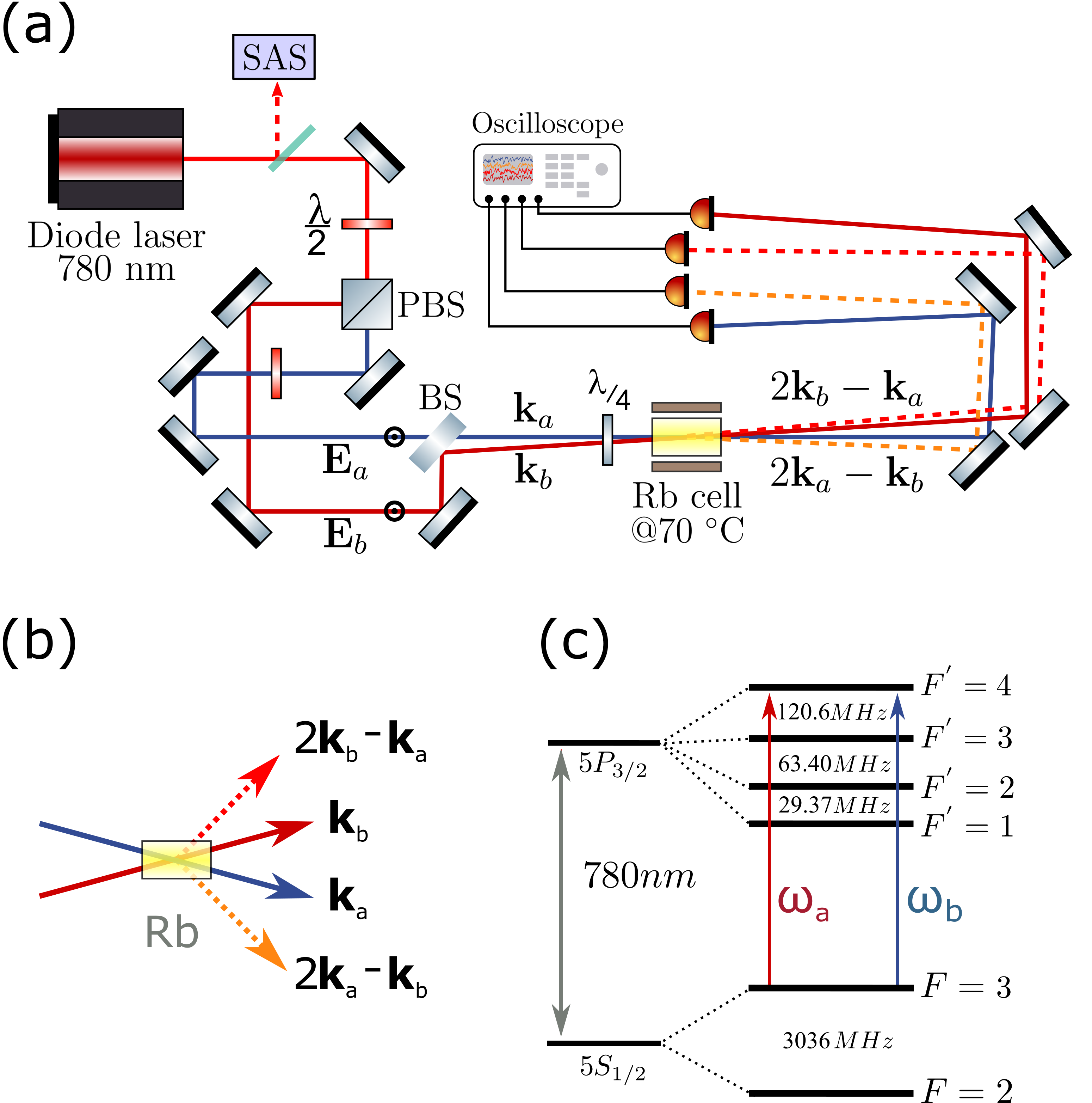}
\caption{Simplified scheme of the experimental setup for the detection of the intensity fluctuations time series of the four participating signals; (b) Wave-vectors of the four signals (two FWM and two transmissions); (c) Hyperfine structure of the $\mathrm{D}_2$ line of $^{85}$Rb.}
\label{fig1}
\end{figure}

We are interested in the two FWM signals generated in directions $2\vec{k}_{a}-\vec{k}_{b}$ and $2\vec{k}_{b}-\vec{k}_{a}$, as shown in Fig. \ref{fig1}(b). Therefore, we investigate processes in which two photons of one of the beams are absorbed and one photon is emitted in the other beam, generating new coherent signals.

The input beams are in an almost copropagating configuration, with a small angle of about 10 mrad between them to allow spatial separation of all four signals. This type of forward geometry, having both input beams with the same intensities and polarizations, is challenging, since scattered light from one beam might arrive at the detection position of the other beams. To avoid this problem, we detect signals far from the cell, exploiting the spatial separation between them. The time series of the intensity fluctuations of the two FWM signals and the transmissions of the input beams $E_a$ and $E_b$ are detected by avalanche photodiodes (APD) with a time resolution of 1 ns. 

The beams that induce the degenerate FWM processes are tuned near the closed transition $\left| F=3 \right\rangle$ $\rightarrow$ $\left| F'=4 \right\rangle$ of the $\mathrm{D}_2$ line of $^{85}$Rb [see Fig. \ref{fig1}(c)]. In fact, we do not lock the frequency of the input laser; instead, we turn off the current modulation and adjust the current so that the laser frequency is located inside the $\left|{F=3}\right\rangle\rightarrow\left|{5^2\mathrm{P}_{3/2}}\right\rangle$ Doppler valley of the saturated absorption spectrum, at a specific detuning from the closed transition.

\begin{figure*}[ht!]
\centering
\includegraphics[width=0.8\linewidth]{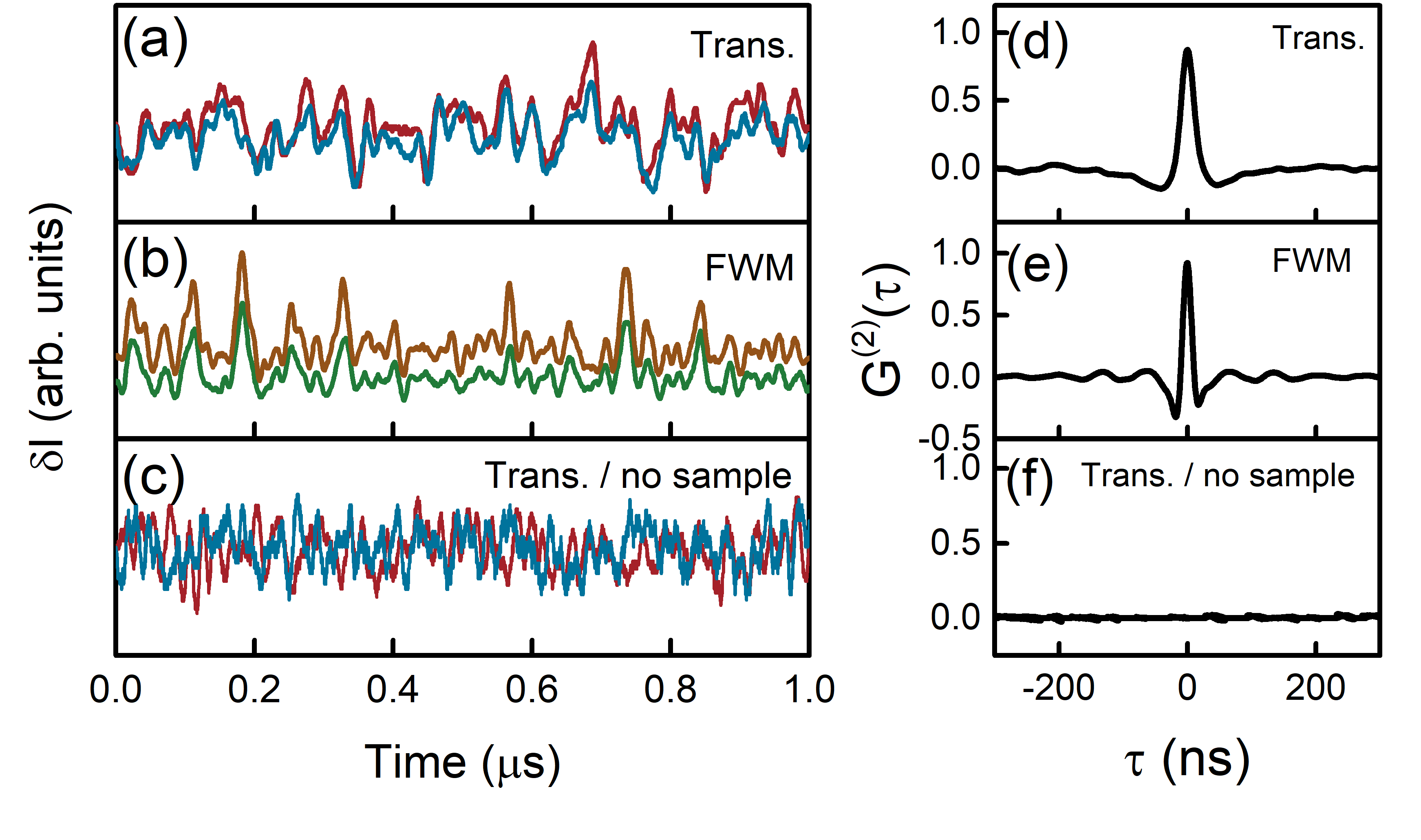}
\caption{Intensity fluctuations time-series for (a) the transmittance of the incident beams with $I_a=I_b=250$ mW/cm$^{2}$ and a detuning from the closed transition of $\delta/2\pi = -250$ MHz, (b) the two symmetric FWM signals generated inside the sample with $I_a=I_b=98$ mW/cm$^{2}$ and  $\delta/2\pi = 90$ MHz, and (c) incident beams in the absence of the Rb cell. To the right, in (d),(e), and (f) we show the cross-correlation curves corresponding to the time-series in (a),(b), and (c), respectively.}
\label{fig2}
\end{figure*}

The time series of the intensity fluctuations of all four signals is presented in Fig. 2. These data are part of the 100 $\mu$s recorded time series and have been filtered with a high-pass ideal FFT filter with a cutoff frequency of 500 kHz to eliminate any slow fluctuations of the signals. In Fig. 2(a), we show the time series for the transmittance of the incident beams (blue and red lines) for an input laser intensity of $I_a=I_b=250$ mW/cm$^{2}$ and a detuning from the closed transition of $\delta/2\pi = -250$ MHz. It is clear that these results are very well synchronized and should present near-perfect correlations, an expected result \cite{ariunbold2010, sautenkov2005, varzhapetyan2009}. Similarly, in Fig. 2(b) we show the intensity fluctuations versus time of the two FWM signals (brown and green lines) for an input laser intensity of $I_a=I_b=98$ mW/cm$^{2}$ and a detuning from the closed transition of $\delta/2\pi = 90$ MHz. It is again noticeable that the fluctuations behave similarly, even though they are not identical.  For comparison, we also present in Fig. 2(c) the time series of the transmission of the incident beams in the absence of the Rb cell. It is clear that in this case, as there is no medium to promote the conversion between phase-to-amplitude-noise, the intensities of the two pumps appear completely uncorrelated.

\begin{figure}
\centering
\includegraphics[width=0.85\linewidth]{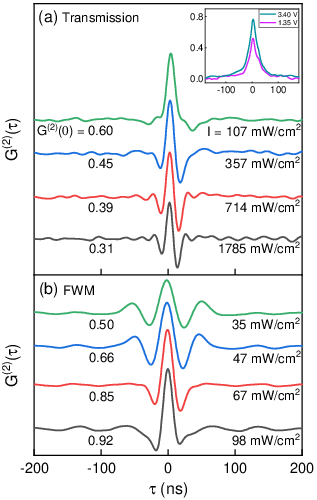}
\caption{ Second-order correlation function $G_{ij}^{(2)}(\tau)$ between (a) the transmitted signals and (b) the FWM signals, for different intensities, $I_a=I_b=I$, at the entrance of the cell. The inset shows cross-correlation curves for the same experimental conditions, $I_a=I_b=71$ mW/cm$^{2}$, with different average signal values at the detector. }
\label{fig3}
\end{figure}

The correlations between the detected signals can be quantified with the second-order correlation function $G_{ij}^{(2)}(\tau)$ \cite{sautenkov2005, varzhapetyan2009, ariunbold2010, yang2012} for the intensity fluctuations of two optical beams with time delay $\tau $. It is given by

\begin{equation}
G_{ij}^{(2)}(\tau)=\frac{\left\langle \delta I_{i}\left(t\right)\delta I_{j}\left(t+\tau\right)\right\rangle }{\sqrt{\left\langle \delta I_{i}\left(t\right)^{2}\right\rangle \left\langle \delta I_{j}\left(t+\tau\right)^{2}\right\rangle }},
\label{g2}
\end{equation}

\noindent
where $\delta I_{i,j} (t) = I_{i,j} (t) - \left\langle I_{i,j} (t)\right\rangle$ are the time-dependent intensity fluctuations with $ \left\langle I_{i,j} (t)\right\rangle$  being the average intensities of the laser fields and $i,j = a, b, s1, s2$  the labels to designate the two input fields and the two FWM signals, respectively. 

\begin{figure}
\centering
\includegraphics[width=1\linewidth]{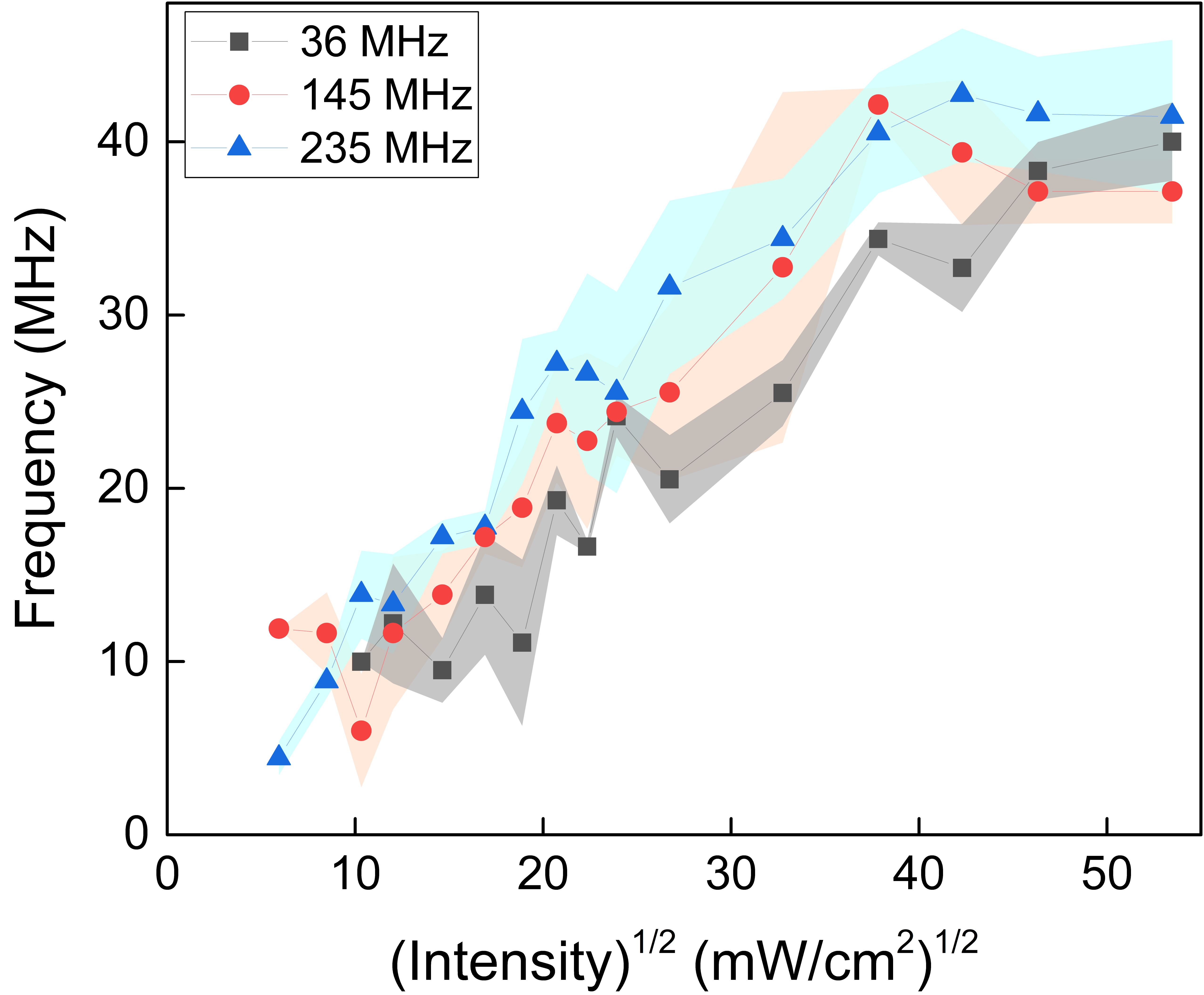}
\caption{Frequency of the oscillation in the correlation function for the transmitted signals as a function of the square root of each beam intensity at the cell's entrance, for three detuning values. The dots are the average values for three measurements, the solid lines are a guide for the eye, and the filled regions indicate the errors.}
\label{fig4}
\end{figure}

We present the intensity fluctuations correlation functions $G_{ij}^{(2)}(\tau)$ on the right side of Fig. 2. Frames (d), (e), and (f) correspond to the pairs of time series of frames (a), (b) and (c), respectively. These correlation functions have peaks at zero time delay with amplitudes (Pearson coefficient) of $\approx 0.87$ for the transmission signals in (a) and more than 0.92 for the FWM signals in (b). This confirms a strong temporal positive correlation in the intensity fluctuations of the output signals. Moreover, a null correlation function for frame (c), without sample, is a clear signature that the cross-correlation that we observe in the transmission beams arises from the resonant phase-noise to amplitude-noise conversion \cite{mcintyre1993, walser1994, camparo1997, camparo1998, camparo1999}. The resonant interaction with the atoms plays a critical role in this result. There would be no correlation if there were no atoms or if the input laser was not near resonance. 

Furthermore, the results indicate that this conversion also occurs with FWM signals, generating correlated fields. This is noteworthy because, although these two FWM signals are excited by the same beams with the same phase fluctuations, they come from processes that cannot occur simultaneously, since a single atom can only generate one of the two FWM signals at a time. As we shall see later, they can also provide information about the temporal dynamics of a set of atoms that interact simultaneously with the same driving fields. The correlation between the fields generated by FWM was observed in a similar experiment using cold atoms when only the stationary group of atoms ($v=0$) contributed to the process. Here, with a system of hot atoms, we need to consider the Maxwell-Boltzmann velocity distribution. In the next section, we build a simple model that considers the velocity distribution and provides information on how the phase fluctuations of the input laser manifest themselves in both transmission and FWM processes.

In Figs. \ref{fig3}(a) and \ref{fig3}(b), we present the second-order correlation function between transmission signals and between FWM signals for different intensities of incident beams at the entrance of the cell. All the results are for circular and parallel polarization. In both figures, the intensity increases from top to bottom. The correlation curves in Fig. 3(a) were obtained from transmission measurements for a fixed detuning of $\delta/2\pi = 235$ MHz and beam diameter of $600$ $\mu$m. In contrast, Fig. 3(b) presents FWM measurements taken for a fixed detuning of $\delta/2\pi = 90$ MHz and a beam diameter of $2.2$ mm. The larger beam diameter is important for obtaining a good FWM signal. It is clear that there are regions of correlation $G_{ij}^{(2)}(\tau)>0$ and regions of anticorrelation $G_{ij}^{(2)}(\tau)<0$, for both signals. Intensity correlation and anticorrelation in the transmitted beams were also reported in Ref. \cite{sautenkov2005}, with a similar experiment using atomic vapors, perpendicularly polarized input beams, and a magnetic field to break the degeneracy of the Zeeman sublevels. This experiment was performed under electromagnetically induced transparency (EIT) conditions, and the authors showed that the transition between correlation and anticorrelation of the fields can be controlled with the magnetic field, changing the detuning of the two-photon resonance. In our experiment, no external magnetic field is introduced, and we work with a two-level system defined by circular and parallel polarized incident beams. Another relevant point concerns the correlation between the FWM signals. Our experimental conditions, defined by the polarization of the input beams, also explain why we observe a positive correlation between these signals and not a competition between them, as described by Yang \textit{et al} \cite{yang2012}.

The most remarkable feature of these results is that the correlation curves indicate an oscillatory behavior for short-time delays ($-100$ ns $ <\tau<100$ ns), with the central peak becoming broader as the input laser intensity decreases. This behavior is clearer in the FWM signal. We also indicate the value of the peaks at zero time delay (Pearson coefficient) for each curve. For the transmission signals, in Fig. 3(a), the Pearson coefficient is quite low for high incident beam intensities and increases as we decrease the input beam intensity. This is expected since, for high intensities, part of the beam passes directly through the sample without interacting with the atoms, so that the phase-modulation to amplitude-modulation conversion process does not occur, resulting in poor intensity correlation. On the other hand, the FWM signal is only generated due to the interaction of the fields with the atomic medium; therefore, the greater the intensity of the laser beams, the greater the intensity of the generated signal and, thus, the Pearson coefficient.

Another feature that might influence the Pearson coefficient is the signal-to-noise ratio(SNR). From an experimental standpoint, one must ensure optimal SNR, as the inset in Fig. 3 indicates. In this graph, we show that for the same conditions, as the average amplitude of the signal decreases using a neutral density filter, the peak value of the correlation decreases as well. We believe that the connection with the SNR relies on the fact that, for a smaller signal, more of the uncorrelated background noise of the detector (an APD) affects the correlation.

In terms of oscillatory behavior, with correlation and anticorrelation regions, it is very similar to what was observed with a sample of cold atoms \cite{almeida2023}. In that previous work, having only atoms with $v=0$ participating in the atom-radiation interaction, it was possible to obtain a detailed map of the correlation as a function of detuning and then verify that these oscillations in the second-order correlation function are connected to the generalized Rabi frequency of the input laser. However, for the present experiment, this last statement needs to be carefully analyzed. First, we indeed have a clear oscillation in the second-order correlation function for both transmission and FWM signals, as shown in Fig. \ref{fig3}. On the other hand, with a hot sample, the detuning cannot be well defined for a fixed laser frequency if we consider all the atomic velocity groups inside the Doppler line. 

A Fourier analysis of the curves in Fig. 3 indicates that they have a spectral component whose frequency is proportional to the intensity. To investigate in detail this behavior, we plotted in Fig. 4, for the transmission signals, the frequency of the oscillation in the correlation as a function of the square root of the input intensity of each beam for three detuning values. The frequency value for each detuning is represented by a different symbol (color). Taking into account the errors in the frequency values (shaded regions), we have a good superposition of the three curves.  It is important to note that as the detuning increases (going away from the Doppler center), the input beams should be absorbed less, so we should expect a higher intensity and, therefore, a higher oscillation frequency, as we observe.

The nearly linear behavior shown in Fig. 4 is clear evidence that these oscillations are connected to the Rabi frequency of the input beams.  This indicates that in the conversion process between phase fluctuations of the laser into intensity fluctuations through the interaction with the atomic medium, the intensity fluctuations oscillate with approximately the generalized Rabi frequency. Similar behavior was described by Papoyan and Shmavonyan when they observed the temporal structure in atomic absorption signal under excitation by a cw phase-diffusion field \cite{papoyan2021}. They argue that this fact is due to the nonadiabatic temporal component of the atom’s response, manifesting itself as population variations oscillating at the Rabi frequency \cite{frueholz1996a}. Furthermore, they also considered that the time of flight could not make a significant contribution and that the effective pulsed excitation is caused by the stochastic phase-fluctuating nature of a cw laser radiation field. In this sense, we also could expect to see this oscillation in our raw data, that is, in the time series of Fig. 2. However, they are not discernible in this case, although we have performed the measurements with a temporal resolution of 1 ns. In fact, a Fourier analysis of those data does not reveal any spectral component in particular.

However, we can make use of a higher-order measurement, which should be able to retrieve the spectral information of the system \cite{chen2010, norris2010}. This is possible with the intensity fluctuations correlation function $G_{ij}^{(2)}(\tau)$, which exhibits a noticeable spectral component, as can be seen in Fig. 3, and confirmed by the Fourier analysis of the transmission signals in Fig. 4. The intriguing feature here is the reason behind the fact that, for a fixed laser frequency, the contribution of different group velocities (inside the Doppler line), with different generalized Rabi frequencies, does not blur these oscillations. As we will see in the next section, only the resonant atoms have a dominant contribution during the transient time, confirming that the nonadiabatic temporal component of the atom’s response is the main mechanism. Moreover, we also confirm that for the present experiment, the effective pulsed excitation is triggered by the time of flight.

\section{\label{sec:level3} Theoretical model}

We consider a phase fluctuation in the incident fields that satisfies a Wiener process, and we employ the model given in Ref. \cite{ariunbold2010} to explain the main features observed in our correlation results. In particular, our treatment to solve the Bloch equations allows us to obtain an independent response for each signal, transmission and FWM. By integrating over the Maxwell-Boltzmann velocity distribution, we can compare our theoretical results with each of the signals and analyze the main contributions.

The treatment of the problem begins considering a two-level system with a total Hamiltonian $\hat{H}=\hat{H}_{0}+\hat{H}_{int}$, where $\hat{H}_{0}$ is the free-atom Hamiltonian and the interaction Hamiltonian is $\hat{H}_{int}=-\hat{\mu}\cdot\vec{E}$, where $\hat{\mu}$ is the electric dipole operator and $\vec{E}=\vec{E}_{a}+\vec{E}_{b}$ is the total electric field. 

The input fields with circular and parallel polarizations are represented by

\begin{equation}
    \begin{split}
        \vec{E}_{a}&=\frac{1}{2}\left[\varepsilon_{a}(t)e^{i\left(\omega_{a}t+\phi\left(t\right)-k_{a}z\right)}\hat{\sigma_{a}}^{+}+c.c.\right];\\\vec{E}_{b}&=\frac{1}{2}\left[\varepsilon_{b}(t)e^{i\left(\omega_{b}t+\phi\left(t\right)-k_{b}z\right)}\hat{\sigma_{b}}^{+}+c.c.\right],
    \end{split}
\end{equation}

\noindent
where $\varepsilon_{l}$ is the amplitude of the electric field; $\omega_{l}$ is the optical frequency, $\phi(t)$ is the fluctuating phase, $\vec{k}_{l}$ is the associated wave-vector and $c.c.$. means the complex conjugate.

We consider that the electric fields have a fluctuating phase $\phi(t)$, described by a Wiener-Levy diffusion process \cite{agarwal76}. For these processes, the average of the stochastic variable is zero and the average of the two-time correlation is given by $\overline{\left\langle \dot{\phi}\left(t\right)\dot{\phi}\left(t'\right)\right\rangle }=2D\delta\left(t-t'\right)$; where $D$ is the diffusion coefficient. Often in the literature, the stochastic process chosen to represent this phase is the Ornstein–Uhlenbeck \cite{gardiner2004quantum, orzag}, which includes an extra term to the Wiener process to make it mean-reversible.

\begin{figure}
\centering
\includegraphics[width=1\linewidth]{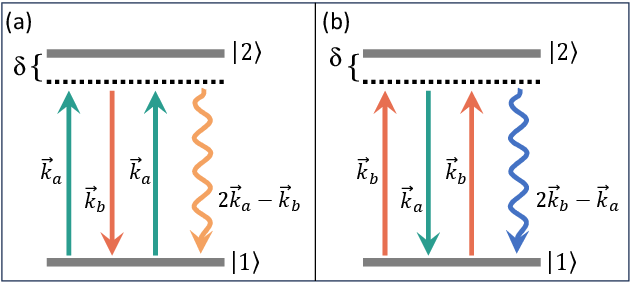}
\caption{Schematic representation of the four-wave mixing parametric processes in a two-level atom generating the FWM signals (a) $(2\mathbf{k}_a-\mathbf{k}_b)$ and (b) $(2\mathbf{k}_b-\mathbf{k}_a)$}.
\label{fig5}
\end{figure}

The nonlinear interaction of the input beams, $E_a$ and $E_b$, with the two-level system, leads to the generation of signal $2\vec{k}_{a}-\vec{k}_{b}$, due to the absorption of two photons from beam $E_a$ and the stimulated emission of one photon from beam $E_b$, and signal $2\vec{k}_{b}-\vec{k}_{a}$, due to the absorption of two photons from $E_b$ and the stimulated emission of one photon from $E_a$, as illustrated schematically in Fig. \ref{fig5}.

The matrix elements of $\hat{H}_{int}$ can be written as

\begin{equation}
H_{int,12}=-\hbar\sum_{l}\Omega_{l}e^{i(\omega_{l}t-k_{l}z)}+c.c.,
\end{equation}

\noindent
where $\Omega_{l}\equiv\frac{\mu_{jk}\varepsilon_{l}}{2\hbar}$ $(l=a$ or $b)$ is the Rabi frequency with $\mu_{jk}$ being the transition dipole moment. The density operator $\hat{\rho}=\sum_{jk}\rho_{jk}\left|j\right\rangle \left\langle k\right|$ describes the state of the atomic ensemble and satisfies $\sum_{j}\rho_{jj}=1$, and $\rho_{jk}=\rho_{kj}^{*}$, where the asterisk means complex conjugation. Its time evolution is given by Liouville’s equation with a relaxation term $\hat{\mathcal{L}}$
associated with spontaneous decay from $\left|2\right\rangle $ to $\left|1\right\rangle $, 

\begin{equation}
\frac{d\hat{\rho}}{dt}=\frac{i}{\hbar}[\hat{\rho},\hat{H}]+\hat{\mathcal{L}}.
\end{equation}

In our model, populations and coherence decay at rates $\Gamma_{jk}$ and $\gamma_{jk}$, respectively. With this, we obtain the optical Bloch equations for the two-level system

\begin{equation}
\begin{split}
\triangle\dot{\rho}&=-\frac{2i}{\hbar}[\rho_{12}H_{int,21}-c.c.]-\Gamma_{21}[\triangle\rho-(\triangle\rho)^{0}],
\\\dot{\rho}_{12}&=-\frac{i}{\hbar}[H_{int,12}\triangle\rho-\rho_{12}\hbar\omega_{21}]-\gamma_{12}\rho_{12},
\end{split}
\end{equation}

\noindent
where $\triangle\rho=(\rho_{22}-\rho_{11})$ is the population difference and $(\triangle\rho)^{0}$
is the population difference far from the region of interaction with fields $E_a$ and $E_b$ and $\omega_{jk}$ is the frequency of the $\left|j\right\rangle$ $\rightarrow$ $\left|k\right\rangle$ transition.

The problem of a two-level system interacting with two strong fields has been addressed in Refs. \cite{agarwal86, friedmann87, wilson-gordon88}, and Eqs. (5) are solved for arbitrary pump intensities assuming that the elements of the density operator oscillate with an infinite number of frequencies, associated with the various Fourier components. In these works, the nonlinear coherence related to FWM processes is found in terms of a recursive formula. Here, we employ a simpler solution method, similar to the treatment found in \cite{boyd81}. We consider three-photon interactions for the coherence describing the FWM process at frequencies $2\omega_{a}-\omega_{b}$ and $2\omega_{b}-\omega_{a}$ and two-photon interactions for the population of the two levels. 
We therefore express the coherence in terms of the Fourier components

\begin{equation}
\begin{split}
\rho_{12}=&\rho_{12}^{\left(\omega_{a}\right)}e^{i\omega_{a}t}+\rho_{12}^{\left(\omega_{b}\right)}e^{i\omega_{b}t}+
\\&\rho_{12}^{\left(2\omega_{a}-\omega_{b}\right)}e^{i\left(2\omega_{a}-\omega_{b}\right)t}+\rho_{12}^{\left(2\omega_{b}-\omega_{a}\right)}e^{i\left(2\omega_{b}-\omega_{a}\right)t}.
\end{split}
\label{Eq.Fourier_rho}
\end{equation}

The two components $\rho_{12}^{\left(2\omega_{a}-\omega_{b}\right)}$ and $\rho_{12}^{\left(2\omega_{b}-\omega_{a}\right)}$ are responsible for the FWM processes that generate the two nonlinear signals. The population difference $\triangle\rho$ has a stationary component and one oscillating at $\left|\omega_{a}-\omega_{b}\right|$

\begin{equation}
\begin{split}
\triangle\rho=&(\triangle\rho)^{dc}+(\triangle\rho)^{\left(\omega_{a}-\omega_{b}\right)}e^{i\left(\omega_{a}-\omega_{b}\right)t}+
\\&(\triangle\rho)^{\left(\omega_{b}-\omega_{a}\right)}e^{-i\left(\omega_{a}-\omega_{b}\right)t}.
\end{split}
\label{Eq.Fourier_population}
\end{equation}

We now substitute Eqs. (3), (6), and (7) into Eqs. (5) and collect terms that oscillate with the same frequency.
The terms for coherence, taking into account the inhomogeneous Doppler broadening, are given by:

\begin{widetext}
\begin{equation}
\begin{aligned}
\dot{\rho}_{12}^{\left(\omega_{a}\right)} = &   \: i\Omega_{a}\left(\Delta\rho\right)^{dc}+i\Omega_{b}(\Delta\rho)^{(\omega_a-\omega_b)}-\left[i\left(\delta_{a}-k_{a}v\right)+\gamma_{12}\right]\rho_{12}^{\left(\omega_{a}\right)}+i\dot{\phi}\rho_{12}^{\left(\omega_{a}\right)};
\\
\dot{\rho}_{12}^{\left(\omega_{b}\right)} = & \: i\Omega_{b}(\Delta\rho)^{dc}+i\Omega_{a}(\Delta\rho)^{(\omega_b-\omega_a)}-\left[i\left(\delta_{b}-k_{b}v\right)+\gamma_{12}\right]\rho_{12}^{\left(\omega_{b}\right)}+i\dot{\phi}\rho_{12}^{\left(\omega_{b}\right)};
\\
\dot{\rho}_{12}^{\left(2\omega_{a}-\omega_{b}\right)}	= &\:i\Omega_{a}(\Delta\rho)^{\left(\omega_{a}-\omega_{b}\right)} - \{i\left[\left(2\delta_{a}-\delta_{b}\right)-\left(2k_{a}-k_{b}\right)v\right] + \gamma_{12}\}\rho_{12}^{\left(2\omega_{a}-\omega_{b}\right)}+i\dot{\phi}\rho_{12}^{\left(2\omega_{a}-\omega_{b}\right)};
\\
\dot{\rho}_{12}^{\left(2\omega_{b}-\omega_{a}\right)}	= &  \: i\Omega_{b}(\Delta\rho)^{\left(\omega_{b}-\omega_{a}\right)} - \{i\left[\left(2\delta_{b}-\delta_{a}\right)-\left(2k_{b} - k_{a}\right)v\right] + \gamma_{12}\}\rho_{12}^{\left(2\omega_{b}-\omega_{a}\right)}+i\dot{\phi}\rho_{12}^{\left(2\omega_{b}-\omega_{a}\right)},
\end{aligned}
\label{Eq.components_coherence}
\end{equation}
\end{widetext}

\noindent
and for population

\begin{widetext}
\begin{equation}
\begin{aligned}
\dot{(\Delta\rho)}^{dc} = &2i[\Omega_{a}^{*}\rho_{12}^{(\omega_{a})}+\Omega_{b}^{*}\rho_{12}^{(\omega_{b})}-\Omega_{a}\rho_{21}^{(\omega_{a})}-\Omega_{b}\rho_{21}^{(\omega_{b})}]-\Gamma_{21}[(\Delta\rho)^{dc}-(\Delta\rho)^{0}];\\
\dot{(\Delta\rho)}^{(\omega_{a}-\omega_{b})} = & 2i[\Omega_{b}^{*}\rho_{12}^{(\omega_{a})} + \Omega_{a}^{*}\rho_{12}^{(2\omega_{a} - \omega_{b})} - \Omega_{a}\rho_{21}^{(\omega_{b})} - \Omega_{b}\rho_{21}^{(2\omega_{b}-\omega_{a})}] - \{i[\left(\delta_{a} - \delta_{b}\right) - \left(k_{a} - k_{b}\right)v] + \Gamma_{21}\} \times(\Delta\rho)^{\left(\omega_{a}-\omega_{b}\right)};\\
\dot{(\Delta\rho)}^{(\omega_{b}-\omega_{a})} = & 2i[\Omega_{a}^{*}\rho_{12}^{(\omega_{b})}+\Omega_{b}^{*}\rho_{12}^{(2\omega_{b}-\omega{a})}-\Omega_{b}\rho_{21}^{(\omega_{a})}-\Omega_{a}\rho_{21}^{(2\omega_{a}-\omega_{b})}] - \{[i\left(\delta_{b}-\delta_{a}\right)-\left(k_{b}-k_{a}\right)v]+\Gamma_{21}]\} \times(\Delta\rho)^{\left(\omega_{b}-\omega_{a}\right)}.
\end{aligned}
\end{equation}
\end{widetext}

\noindent
where $\delta_{l}\equiv\omega_{l}-\omega_{21}$ and $k_{l}$ are the detuning and wave number of field $E_l$. As indicated by the set of Eqs. (8), this treatment allows us to separate the response of each signal, transmission and FWM, which are described by the $\rho_{12}$ coherence. 

Since the set of Eqs. (8) contains stochastical terms, we must solve them numerically using Itô's calculus. As previously mentioned, we take a typical stationary stochastic process, the Ornstein-Uhlenbeck process, to describe the phase fluctuations. This process satisfies the SDE:

\begin{equation}
    dX_{t}=\alpha\left(\gamma-X_{t}\right)dt+\beta dW_{t}
\end{equation}

\noindent
where the Itô's diffusive process $dX_{t}$ has a deterministic part and a stochastic one. The deterministic term, the first one, has a magnitude of the mean drift $\alpha$ while the asymptotic mean is $\gamma$. If $X_t>\gamma$ the drift will be negative and the process will go towards the mean. If $X_t<\gamma$ then the opposite happens, the drift is positive and the process moves away from the mean. As for the stochastic part, it is a Brownian motion $W_t$ with a magnitude constant $\beta$.

We solve the system of SDEs using a stochastic Runge-Kutta for scalar noise algorithm. This algorithm possesses good accuracy for our problem, with a thin distribution of residuals. We also use the same Brownian increment $dW_t$ for both one-photon coherences, as the original fluctuation comes from a single laser. Finally, we probed several choices of parameters of the Ornstein-Uhlenbeck process, but the outcomes are not drastically different as long as the variance of the process, given by $\beta^2 / 2\alpha$, is small.

A numerical simulation is performed for each velocity value and then added following the Maxwell-Boltzmann velocity distribution. The addition is done at each instant of time, and the numerical calculations for the set of velocities use the same noise seed. Once this step is completed, we have access to the theoretical time series of all elements of the density matrix, taking into account the contribution of every atom in the cell. Therefore, for each of these terms, we can calculate the second-order correlation function for intensity fluctuations. However, we must establish the link between the density-matrix elements and the actual detected signal. To do so, we solve the wave equation derived from Maxwell's equations, neglecting the transverse derivatives of the electric field. 

With a few algebraic manipulations and using the adiabatic and nondepleted beam approximations, we can obtain decoupled differential equations for transmitted beams $\frac{\partial\Omega_{l}}{\partial z}=i\kappa_{12}\rho_{12}^{(\omega_{l})}$, and for the FWM generated beams $\frac{\partial\Omega_{FWM}^{(r)}}{\partial z}=i\kappa_{12}\rho_{12}^{(r)}$, where $l=a$ or $b$ , $r=2\omega_{a}-\omega_{b}$ or $2\omega_{b}-\omega_{a}$, and $\kappa_{12}=\frac{\omega_{l}N\mu_{12}^{2}}{2\hbar\epsilon_{0}c}$, with $N$ being the number of atoms. Solving these equations leads to the fields we detect in the experiment after they propagate in the sample. To do this, we use the fact that the cell length $L$ is smaller than the Rayleigh length of the fields at play. Therefore, it is adequate to consider the thin-medium regime, which implies that we can make use of the equations in Ref. \cite{ariunbold2010} and rewrite the second-order correlation function of the intensity fluctuations of Eq. (1) for transmissions as:

\begin{equation}
    G^{\left(2\right)}\left(\tau\right)=\frac{\langle \mathrm{Im}(\delta\rho_{12}^{(\omega_{a})}\left(t\right))\mathrm{Im}(\delta\rho_{12}^{(\omega_{b})}\left(t+\tau\right))\rangle }{\sqrt{\langle [\mathrm{Im}(\delta\rho_{12}^{(\omega_{a})}\left(t\right))]^{2}\rangle \langle [\mathrm{Im}(\delta\rho_{12}^{(\omega_{b})}\left(t\right))]^{2}\rangle }}.
\end{equation}

\noindent
and for the FWM signals as:

\begin{equation}
G^{\left(2\right)}\left(\tau\right)=\frac{\langle \mathrm{Im}(\delta\rho_{12}^{(2\omega_{a}-\omega_{b})}\left(t\right))\mathrm{Im}(\delta\rho_{12}^{(2\omega_{b}-\omega_{a})}\left(t+\tau\right))\rangle }{\sqrt{\langle[\mathrm{Im}(\delta\rho_{12}^{(2\omega_{a}-\omega_{b})}\left(t\right))]^{2}\rangle \langle [\mathrm{Im}(\delta\rho_{12}^{(2\omega_{b}-\omega_{a})}\left(t\right))]^{2}\rangle }}.
\end{equation}




\noindent
To obtain this last result, we neglected second-order terms when calculating the field intensity.

A similar theoretical analysis of the correlation between field intensity fluctuations of two FWM signals generated in a two-level system is presented in Ref. \cite{miranda2024}. In that work, the authors focused on the problem of a sample of cold atoms ($v=0$) and studied the behavior of the correlation function $G_{ij}^{(2)}(\tau)$ as the detuning of the driven fields varies. However, our experiments were performed in an atomic vapor, so the velocity distribution needs to be taken into account. Another important aspect is that we measure both the transmit signal and the FWM signal, so it is interesting to compare both. This is discussed in the next section, where we present our theoretical results.

\section{\label{sec:level4}Theoretical results}

After solving the coupled SDE system, we have a numerical simulation of the time series for the transmission and FWM signals. An example of a single realization of such a series for both signals, transmission (green) and FWM (red), is presented in Fig. 6 (a) for a detuning of $\delta/2 \pi = 30$ MHz, a Rabi frequency $\Omega_a=\Omega_b= \Gamma$ and taking into account the velocity groups $-0.2$u $<v<0.2$u, where u is the most probable velocity. In the first 1 $\mu$s of the simulated signal, we draw attention to the transient regime, where both series present an oscillatory behavior, with an abrupt variation of the transmission signal due to the term of population difference dc. We also show, in the inset, a zoom of the time series (for the transmission signal) away from the transient response. Employing Eq. (13), we calculate the second-order correlation function for the series of Fig. 6(a) and present the results in Fig. 6(b). Since we use the same Brownian increment $dW_t$ for both signals, they must be perfectly correlated, as in Fig. 6(b). It is easy to introduce different increments for each input field and control how large their correlation is by stating that $dW_t^{(1)}=dW_t^{(2)}+\sqrt{1-\rho^2}dW_t^{(3)}$, where $\rho$ ranges from zero to one; that is, one increment is equal to the other with the addition of a third increment. However, we use a single cw laser, so the stochastic phase each input field carries should be the same. The result shown in Fig. 6(b) can be compared with that obtained from the experiment as shown in Fig. 3(a) for the transmitted beams and in Fig. 3(b) for the FWM signal. Although in the experiment it is difficult to obtain a good correlation in the same experimental conditions in both signals, this is not true for the auto-correlations calculated from the theoretical time series. As shown in Fig. 6(b) both correlations present an oscillatory behavior, close to $\tau = 0$, very similar to that observed in the experiment. The existence of a difference in the oscillation frequency between the two signals is not very clear, and a careful investigation, including an experimental verification, is in progress. It is interesting to note that if we neglect the transient regime, the shape of the peak in the second-order correlation function changes, it becomes similar to a delta function (see inset), which indicates the presence of only white noise. 

\begin{figure}
\centering
\includegraphics[width=1\linewidth]{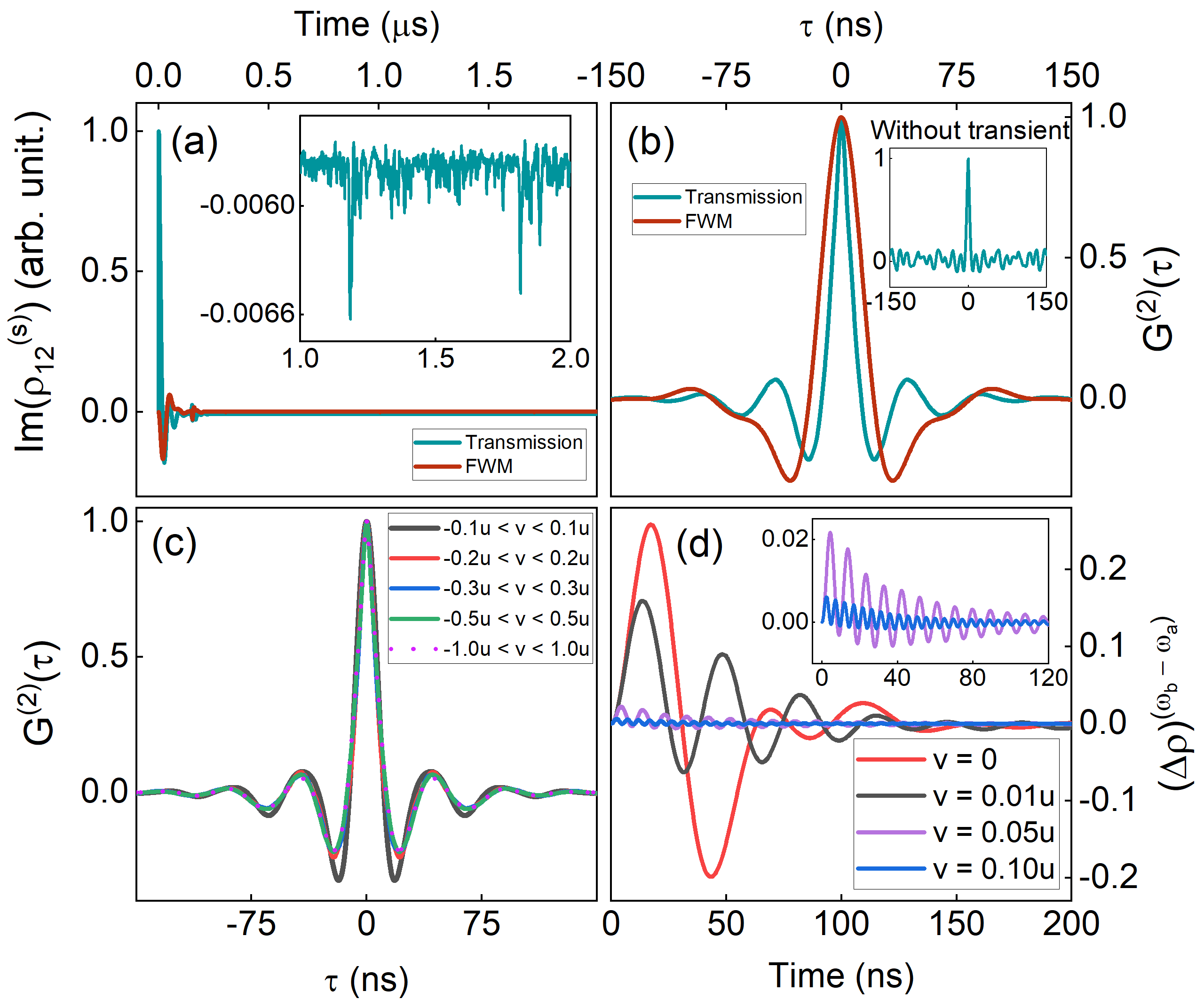}
\caption{(a) Numerical simulation of a time series of the intensity fluctuations for the transmission $Im(\rho_{12}^{(s=\omega_{a})})$ (green) and FWM $Im(\rho_{12}^{(s=2\omega_{a}-\omega_{b})})$  (red) signals with input Rabi frequency $\Omega_a=\Omega_b = \Gamma$, detuning from the excited state of $\delta/2 \pi = 30$ MHz and taking into account the velocity groups $-0.2$u $<v<0.2$u. (b) Cross-correlation between transmitted signals (green) and FWM signals (red), for the time series of (a). The inset shows the cross-correlation when the transient is not taken into account. (c) Cross-correlation between transmitted signals considering the contribution of different velocity groups. (d) Transient dynamics of $(\Delta\rho)^{(\omega_{b}-\omega_{a})}$ for different velocity groups. The inset is the zoom for $v=0.05$u and $v=0.1$u.}
\label{fig6}
\end{figure}

The contribution of velocity distribution in a vapor can be better understood if we analyze the second-order correlation function for different velocity groups, as shown in Fig. 6(c). We see almost no variation in the second-order correlation function when we consider the contribution of velocity groups in the range greater than $-0.2$u $<v<0.2$u. This result is an indication that only the atoms close to the resonance make a real contribution to the central peak of $G^{(2)}(\tau)$. Another important feature is the effect of the transient dynamics of the atomic response. Fig. 6(d) shows how $(\Delta\rho)^{(\omega_{b}-\omega_{a})}$ varies during the transient regime for different velocity groups. We see that the Rabi oscillation of atoms with a velocity equal to or greater than $0.1$u is smaller than one order of magnitude compared to atoms at resonance. These results confirm that only atoms at or near resonance are actually responsible for the shape of the central peak of $G^{(2)}(\tau)$ and that the observed oscillation near $\tau=0$ is connected to the Rabi frequency. 

\begin{figure}
\centering
\includegraphics[width=1\linewidth]{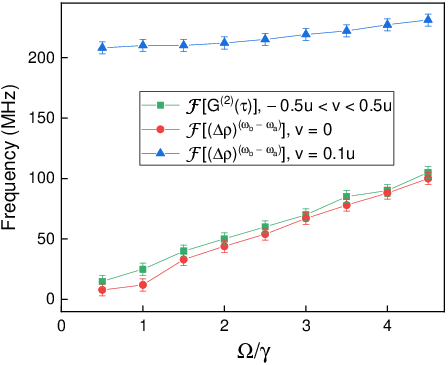}
\caption{Dominant oscillation frequency of the cross-correlation function between transmitted signals considering velocity groups $-0.5$u $<v<0.5$u (green squares), and of the $(\Delta\rho)^{(\omega_{b}-\omega_{a})}$ for velocity groups $v=0$ (red circles) and $v=0.1$u (blue triangles), as a function of the pump Rabi frequency $\Omega/\gamma$.}
\label{fig7}
\end{figure}

A graph similar to the one shown in Fig. 4, for the dominant oscillation frequency of the cross-correlation function, $\mathcal{F}[G^{(2)}(\tau)]$, is presented in Fig. 7 for the transmission signals, considering velocity groups $-0.5$u $<v<0.5$u (green squares). We also plotted $\mathcal{F}[(\Delta\rho)^{(\omega_{b}-\omega_{a})}]$ in the velocity groups $v=0$ (red circles) and $v=0.1$u (blue triangles), as a function of the pump Rabi frequency $\Omega/\gamma$. The results confirm that the frequency of oscillation that we see in the correlation curves has the largest contribution from the resonant atoms. Furthermore, they argue that this oscillation frequency in $G^{(2)}(\tau)$ is indeed well described by the generalized Rabi frequency.

It is important to reinforce that this oscillating behavior has already been observed in experiments carried out in a cloud of cold atoms \cite{almeida2023}. However, in a vapor, we would expect this signature to be erased because of atomic motion. One could expect that the Doppler integration would blur these oscillations since the correlation curve would contain the response of several velocity groups. In this sense, to observe a clear signature of oscillatory behavior, some aspects are really fundamental. First, as we have shown, the transient regime is essential, and therefore a short initial excitation pulse is required. In our system, the effective pulse excitation is determined by the flight of atoms within the laser beam. For a mean thermal velocity of Rb atoms u$\approx2.5\times10^{4}$ cm/s, the flight of an atom into the light beam corresponds to a pulse with a duration of a few $\mu$s. Another fact is related to our detection resolution, of the order of 1 ns, determined by the APDs. However, these two points are not sufficient to observe the Rabi oscillation directly from the time series, as reported in Ref. \cite{papoyan2021}. We need to go to a higher-order measurement to be able to recover the spectral information of the system. In particular, we use the intensity-fluctuation correlation function $G^{(2)}(\tau)$, which presents a noticeable spectral component, as indicated by the experimental results and confirmed by our theoretical treatment.

\section{\label{sec:level5}Conclusions}
Our study investigates the influence of the atomic velocity distribution on the correlation between intensity fluctuations of the transmission and FWM signals in hot rubidium vapor. We present experimental and theoretical results that demonstrate an oscillatory behavior in the correlations, strongly dependent on the intensity of the incident fields. This behavior is a clear signature of Rabi oscillations, which are preserved even in the presence of the Doppler broadening. 

Notably, we observe that this oscillatory behavior in the second-order correlation functions is more pronounced in the FWM signals. This highlights the nonlinear process's ability to retain spectral information about the atom-field interaction, even in an inhomogeneously broadened sample. The results also suggest a dominant contribution from specific velocity groups, emphasizing the relevance of transient excitation and detection resolution in resolving these oscillatory features.

Our theoretical model, based on stochastic differential equations and incorporating the Maxwell-Boltzmann velocity distribution, agrees well with the experimental data. One of the model's achievements is showing that the near-resonant group velocities are the main cause of the correlation and, consequently, the Rabi oscillation we measure. Moreover, the model separates the transmission from the FWM signal, although we believe more data is needed to confirm its agreement with the nonlinear signal.

\begin{acknowledgments}
This work was supported by Coordena\c{c}\~{a}o de Aperfei\c{c}oamento de Pessoal de N\'{\i}vel Superior (CAPES - PROEX Grant 534/2018, No.
23038.003382/2018-39) and Funda\c{c}\~{a}o de Amparo \`{a} Pesquisa do Estado de S\~{a}o Paulo (FAPESP - Grant 2021/06535-0). A. A. C. de Almeida acknowledges financial support from Conselho Nacional de Desenvolvimento Cient\'{\i}fico e Tecnol\'{o}gico (CNPq - Grant
351985/2023-9) and Office of Naval Research (ONR - Grant N62909-23-1-2014). A. S. Alvarez acknowledges financial support from CAPES (88887.473379/2020-00). M. R. L. da Motta acknowledges financial support from CAPES (Grant 88887.623521/2021-00). S. S. Vianna acknowledges financial support from CNPq (Grant 307722/2023-6).
\end{acknowledgments}

\bibliography{biblio}

\end{document}